\documentclass[]{spie}  

\usepackage{amsmath,amsfonts,amssymb}
\usepackage{graphicx}
\usepackage[colorlinks=true, allcolors=blue]{hyperref}
\usepackage{multirow}
\usepackage{array}
\usepackage{longtable}
\usepackage{soul}
\usepackage{enumitem}

\usepackage{xcolor,colortbl}
\usepackage[]{aas_macros}

\graphicspath{{figures/}}

\title{On-sky demonstration of self-learning predictive control with MagAO-X}

\author{
Sebastiaan Y. Haffert$^{*,a,b}$, J.R. Males$^{b}$, P.T. Johnson$^{b}$, L.M. Close$^{b}$, O. Guyon$^{b}$, J. Kueny$^{b}$, J. Liberman$^{b}$ , J.D. Long$^{c}$, M. Lucas$^{b}$, E. McEwen$^{b}$, T. Nguyen$^{b}$, A. Taras$^{a}$, K. Van Gorkom$^{b}$, M. Kautz$^{b}$, K. Twitchell$^{b}$, L. Schatz$^{d}$
\\
\vspace{0.2cm}
$^{a}$Leiden University, Einsteinweg 55, 2233 CC, Leiden, The Netherlands\\
$^{b}$Steward Observatory, 933 North Cherry Avenue, Tucson, AZ 85721-0065, USA\\
$^{c}$Center for Computational Astrophysics, Flatiron Institute, 162 5th Avenue, New York, New
York\\
$^{d}$Starfire Optical Range, Kirtland Air Force Base, Albuquerque, New Mexico}

\authorinfo{*sebastiaan.haffert@strw.leidenuniv.nl}

\begin{document} 
\maketitle


\begin{abstract}Direct imaging of exoplanets is very tricky and requires extremely well corrected wavefronts. Especially low-order order modes are detrimental to the performance of coronagraphs at their inner-working angle. However, that is precisely where conventional AO systems have the highest residuals that are caused by servo-lag errors. This servo-lag error can be reduced with predictive control where the control anticipates the future state of the atmospheric disturbance. We use a self-learning model predictive controller based on the concepts from sub-space predictive control (SPC). We present a novel implementation of the SPC by using an online QR-decomposition based recursive least squares approach. This approach has now been used for self-learning control of vibrations on the MagAO-X instrument. We see on average an Strehl increase of 15 percent and a decrease of the jitter to 0.9 mas. I will discuss how we have implemented the controller and its on-sky perfomance.\end{abstract}



\section{Introduction}

The direct imaging and characterization of exoplanets requires extreme control of optical aberrations to suppress the overwhelming brightness of the host star and reveal the faint planetary signal \cite{Kenworthy2025HighContrast}. Coronagraphic instruments for current and future high-contrast imaging facilities, including the Giant Magellan Telescope and future space-based missions, require wavefront errors to be controlled at unprecedented levels \cite{2018ARA&A..56..315G}. In particular, residual low-order aberrations, such as tip/tilt errors, are especially detrimental because they directly couple stellar light into the dark region around the coronagraphic inner working angle. Therefore, improving the correction of these modes is critical for maximizing the scientific performance of high-contrast imaging systems.

Adaptive optics (AO) systems have traditionally relied on feedback control approaches that correct the measured wavefront errors based on the current estimate of the atmospheric disturbance. However, the finite latency between wavefront sensing, computation, and deformable mirror actuation introduces a temporal delay that limits the achievable correction\cite{Poyneer2007}. This servo-lag error is one of the dominant residual error terms in extreme adaptive optics (ExAO) systems, particularly at high spatial frequencies and for rapidly varying atmospheric conditions\cite{2018ARA&A..56..315G,Kenworthy2025HighContrast}. Conventional integrator controllers are inherently reactive and cannot compensate for future atmospheric evolution, motivating the development of predictive control approaches that anticipate the incoming wavefront disturbance\cite{Poyneer2007,Petit2008,Males2018}.

Predictive control methods use temporal models of the wavefront evolution to estimate and compensate for future disturbances before they occur. A promising approach is sub-space predictive control (SPC)\cite{huang2008dynamic}, which uses data-driven system identification to construct a model directly from wavefront sensor telemetry without requiring explicit assumptions about atmospheric turbulence statistics. SPC has previously demonstrated significant improvements in adaptive optics simulations and laboratory experiments \cite{haffert2021data}, showing the potential of data-driven predictive control for next-generation extreme adaptive optics systems. However, practical implementation on astronomical instruments requires computationally efficient and numerically robust algorithms capable of continuously adapting to changing observing conditions\cite{haffert2021lab}.

In this work, we present a new implementation of SPC based on an online recursive least-squares algorithm using QR decomposition\cite{McWhirter1982,Haykin2002}. This formulation improves the numerical stability of the adaptive model estimation and enables robust real-time operation on an astronomical instrument. We have implemented this predictive controller on the MagAO-X extreme adaptive optics instrument\cite{males_2024} and applied it to the correction of low-order tip/tilt disturbances. Using on-sky telemetry, we demonstrate improved wavefront correction performance, achieving an average Strehl ratio increase of 15\% and reducing residual image jitter to 0.9 mas. We discuss the controller architecture, the real-time implementation, and the on-sky performance of predictive control for high-contrast imaging.

\section{Data-driven Sub-space Predictive Control}

Predictive control provides a powerful approach to mitigate servo-lag errors in adaptive optics by anticipating the future evolution of the residual wavefront. Unlike conventional feedback controllers, which react to measured residual errors, predictive controllers use a temporal model of the adaptive optics system to determine the control action that minimizes future residual aberrations. In this work, we use a data-driven implementation of Sub-space Predictive Control (SPC), originally introduced by Haffert et al.~\cite{Haffert2021SPC}. The key concept of SPC is to identify the closed-loop input-output behavior of the adaptive optics system directly from telemetry data, avoiding the need for explicit assumptions about atmospheric turbulence statistics or instrument dynamics.

The SPC approach models the evolution of future residual wavefront errors as a linear function of measured residual errors and applied control commands. The residual wavefront at future time steps is predicted according to

\begin{equation}
\mathbf{e}_{f}
=
\begin{bmatrix}
\mathbf{A} & \mathbf{B} & \mathbf{C}
\end{bmatrix}
\begin{bmatrix}
\mathbf{e}_{p}\\
\Delta \mathbf{u}_{p}\\
\Delta \mathbf{u}_{f}
\end{bmatrix},
\end{equation}

where $\mathbf{e}_{f}$ represents the vector of predicted future residual wavefront errors, $\mathbf{e}_{p}$ contains the history of measured residual errors, $\Delta\mathbf{u}_{p}$ contains the history of applied control increments, and $\Delta\mathbf{u}_{f}$ represents the future control increments. The matrices $\mathbf{A}$, $\mathbf{B}$, and $\mathbf{C}$ describe the learned closed-loop dynamics of the adaptive optics system.

A key feature of this formulation is that the future control increments are explicitly included in the prediction model. Therefore, the model does not simply extrapolate the atmospheric disturbance; instead, it predicts how the residual wavefront will evolve under a proposed future control sequence. The optimal control command is obtained by minimizing the predicted future residual error:

\begin{equation}
\Delta\mathbf{u}_{f}^{*}
=
\underset{\Delta\mathbf{u}_{f}}{\mathrm{argmin}}
\left\|
\mathbf{e}_{f}
\right\|^{2}.
\end{equation}

The first element of the optimal future control sequence is then applied to the deformable mirror command:

\begin{equation}
\mathbf{u}_{k+1}
=
\mathbf{u}_{k}
+
\Delta\mathbf{u}_{k}^{*}.
\end{equation}

This receding horizon approach allows the controller to continuously update its prediction based on new wavefront sensor measurements while compensating for the temporal delay inherent in adaptive optics systems.

The model parameters are learned online from telemetry using Recursive Least Squares (RLS). In the original SPC implementation, RLS was used to estimate the matrices $\mathbf{A}$, $\mathbf{B}$, and $\mathbf{C}$ from measured residual errors and applied control commands. Although computationally efficient, the conventional covariance-form RLS algorithm can suffer from numerical conditioning problems during long-duration operation, especially when the input vectors contain strongly correlated measurements as is typical for adaptive optics telemetry.

In this work, we introduce a QR-decomposition-based Recursive Least Squares (QRD-RLS) implementation for online model identification. Instead of explicitly updating and inverting the covariance matrix, QRD-RLS maintains a numerically stable QR factorization of the least-squares problem. This improves robustness during continuous adaptation and enables reliable real-time operation on astronomical instrumentation.

We apply the QRD-RLS SPC framework to low-order wavefront control on the MagAO-X extreme adaptive optics instrument. While the SPC framework is general and can be extended to high-order wavefront control, we first focus on tip/tilt correction because these modes strongly affect coronagraphic performance and are dominated by temporal servo-lag errors. This implementation demonstrates the potential of adaptive data-driven predictive control for improving the performance of high-contrast imaging systems.

\subsection{Hybrid Predictive-Integral Control}

Although the data-driven SPC controller provides optimal predictive control when the identified model accurately describes the system dynamics, the learned model is initially uncertain during the start-up phase. In addition, changes in atmospheric conditions or instrument dynamics can temporarily reduce the accuracy of the learned model. To ensure robust operation throughout the entire learning process, we combine the predictive controller with a conventional integral controller.

The final control update is composed of two contributions:

\begin{equation}
    \mathbf{u}_{k+1}
    =
    \mathbf{u}_{k}
    +
    \Delta\mathbf{u}_{k}^{\mathrm{SPC}}
    +
    \Delta\mathbf{u}_{k}^{\mathrm{I}},
\end{equation}

where $\Delta\mathbf{u}_{k}^{\mathrm{SPC}}$ is the optimal control increment obtained by minimizing the predicted future residuals, and $\Delta\mathbf{u}_{k}^{\mathrm{I}}$ is the contribution from the integral controller. The integral control term is given by

\begin{equation}
    \Delta\mathbf{u}_{k}^{\mathrm{I}}
    =
    g_{\mathrm{I}}\mathbf{e}_{k},
\end{equation}

where $g_{\mathrm{I}}$ is the integral gain and $\mathbf{e}_{k}$ is the measured residual wavefront error.

The integral controller provides a stable fallback mechanism during periods where the predictive model is not yet sufficiently identified. This is particularly important during the initial acquisition of telemetry, where the matrices describing the closed-loop dynamics are poorly constrained. As the SPC model continuously updates from the measured residuals and applied commands, the predictive controller gradually learns the dynamics that generate the integral control response.

Because the integral controller is part of the closed-loop system used for model identification, the learned SPC model automatically incorporates the static correction component provided by the integrator. Consequently, the predictive controller progressively takes over the correction of predictable disturbances, while the integral controller primarily compensates for non-modeled effects and slow drifts. This hybrid approach combines the robustness of classical integral control with the improved temporal correction capability of predictive control.

\section{On-sky Demonstration of Vibration Predictive Control}

\begin{figure}[ht]
\centering
\includegraphics[width=0.75\textwidth]{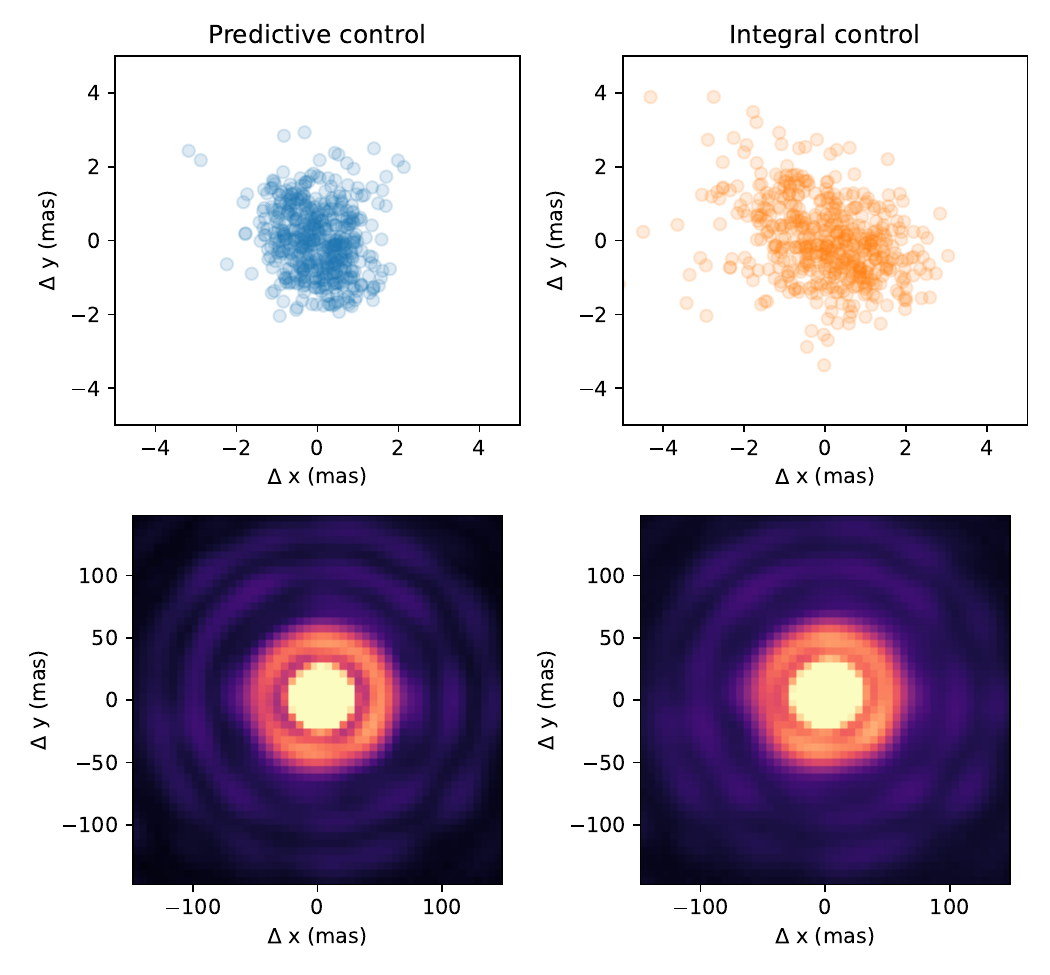}
\caption{The left column shows the results from the predictive controller and the right column shows the optimized modal gain controller. The bottom panels show the PSF at 908 nm on a square-root scaling. The colormap is scaled from 0 to 0.1 of the PSF peak. This colormap scaling highlights the PSF structure. The top panels show the measured PSF centroids sampled at 22 Hz.}
\label{fig:fiber}
\end{figure}

We implemented the QRD-RLS-based data-driven sub-space predictive control (DDSPC) approach on the MagAO-X instrument for on-sky vibration predictive control. In this demonstration, the controller was applied to the low-order tip/tilt modes, while allowing a full modal cross-talk matrix in the predictive model. This enables the controller to learn coupled temporal behavior between modes rather than independently predicting each mode separately. The results of the on-sky experiment are shown in Figure \ref{fig:fiber}.

During the experiment, the science camera operated at 22~Hz, while the wavefront sensor provided measurements at 2~kHz. The DDSPC model used a history of 10 previous residual measurements and control commands to predict a future horizon of 3 samples. The controller was initialized and trained online for 3000 iterations, corresponding to approximately 1.5~seconds of on-sky telemetry. After this initial learning phase, the learned QRD-RLS predictive controller was used in combination with the integral controller to compensate for predictable temporal disturbances.

\begin{figure}[ht]
\centering
\includegraphics[width=0.75\textwidth]{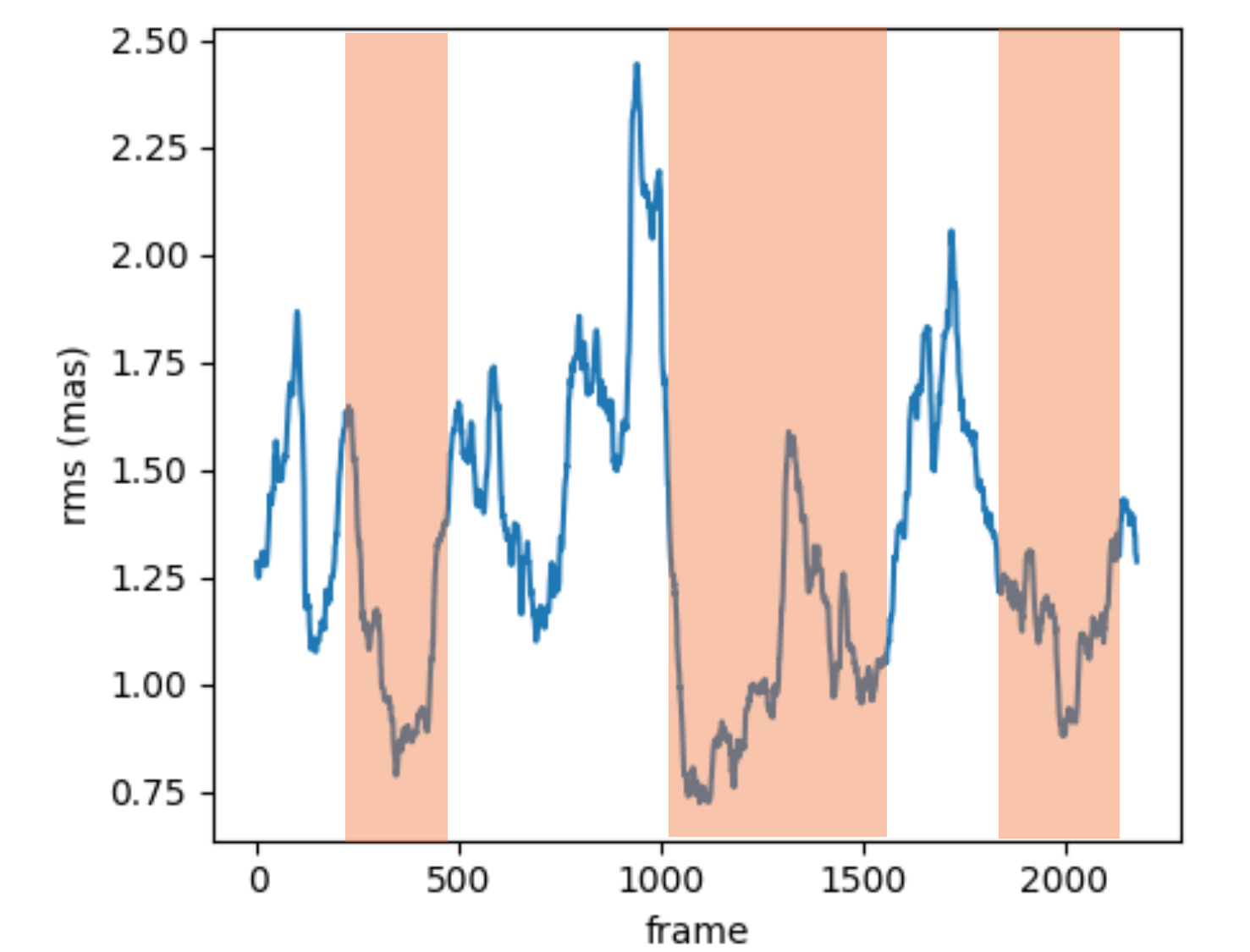}
\caption{Time series of several on/off experiments where we turned the predictive control on and off to test repeatability. The orange shaded areas correspond roughly to when the predictive control was turned on.}
\label{fig:onoff}
\end{figure}

We compare the on-sky point spread functions obtained with the QRD-RLS DDSPC controller to a conventional integral controller. Using a predictive model with a history of 10 frames, the measured image jitter is reduced from 1.29~mas to 0.85~mas, corresponding to a 34\% reduction in residual jitter. The improved temporal correction results in a 15.3\% increase in measured Strehl ratio. The measured jitter reduction accounts for part of the observed Strehl improvement; the remaining improvement is consistent with additional suppression of higher-frequency temporal disturbances beyond the 22~Hz bandwidth of the science camera measurements. A detailed analysis of the contribution from these higher-frequency components, including direct wavefront telemetry analysis, will be presented in future work.

To evaluate the repeatability of the predictive controller, we performed a series of on-sky experiments in which the QRD-RLS DDSPC controller was repeatedly enabled and disabled while observing under similar conditions. Figure \ref{fig:onoff} shows the resulting time series of the measured performance metrics. The orange shaded regions indicate the periods during which predictive control was enabled. Each activation of the predictive controller resulted in an immediate improvement in image quality, while disabling the controller caused the performance to return to the baseline level achieved by the integral controller alone. The consistent improvement over multiple on/off cycles demonstrates that the observed gain is a direct consequence of the predictive controller rather than changes in the atmospheric conditions or observing environment, highlighting the robustness and repeatability of the proposed approach during on-sky operation.

\section{Conclusion}

We have presented a numerically robust implementation of data-driven Sub-space Predictive Control based on a QR-decomposition Recursive Least Squares (QRD-RLS) algorithm. The new formulation enables stable online identification of the predictive model while maintaining the adaptive, self-learning characteristics of the original SPC framework. By combining the predictive controller with a conventional integral controller, the system remains robust during model initialization while progressively learning to compensate predictable disturbances directly from on-sky telemetry.

The controller was successfully implemented on the MagAO-X extreme adaptive optics instrument and demonstrated on-sky for low-order tip/tilt vibration control. The predictive controller consistently reduced the residual image jitter from 1.29~mas to 0.85~mas while increasing the measured Strehl ratio by 15.3\%. Repeated on/off experiments confirmed that these improvements are reproducible and directly attributable to the predictive controller, demonstrating the maturity of the approach for operation on astronomical adaptive optics systems.

The next stage of this work is to extend the controller beyond tip/tilt and apply predictive control to progressively larger sets of wavefront modes, ultimately targeting the full $\sim$1600-mode correction space of MagAO-X. This expanded implementation is planned for on-sky testing during the November 2026 MagAO-X observing run and represents an important step toward predictive control for next-generation extreme adaptive optics systems operating at the spatial and temporal limits required for high-contrast exoplanet imaging.

The next stage of this work is to extend the controller beyond tip/tilt and apply predictive control to progressively larger sets of wavefront modes, ultimately targeting the full $\sim$1600-mode correction space of MagAO-X. This expanded implementation is planned for on-sky testing during the November 2026 MagAO-X observing run and represents an important step toward predictive control for next-generation extreme adaptive optics systems operating at the spatial and temporal limits required for high-contrast exoplanet imaging.

Future work will incorporate sensor fusion by combining wavefront sensor telemetry with high signal-to-noise accelerometer measurements acquired at up to 8~kHz from accelerometers mounted on the telescope top ring. Recent measurements on MagAO-X have shown that approximately one-third of the residual tip/tilt wavefront error is coherent with structural vibrations and that these vibrations can be predicted from accelerometer telemetry \cite{Johnson2026Vibrations}. These measurements provide a complementary, high-bandwidth sensing channel that is unavailable from the wavefront sensor alone. By combining the 8~kHz accelerometer telemetry with the 2~kHz wavefront sensor measurements within the data-driven SPC framework, we aim to develop a multi-rate sensor-fusion predictive controller capable of simultaneously mitigating atmospheric servo-lag and structural vibrations. This capability will be particularly valuable for future facilities such as GMagAO-X \cite{Males2026GMagAOX,Close2026GMagAOXFDR,Haffert2026GMagAOX} and the Planetary Camera and Spectrograph (PCS)\cite{kasper2021pcs}, where sub-nanometer wavefront stability over thousands of controlled modes will be required for high-contrast exoplanet imaging \cite{Haffert2026GMagAOX}.

\acknowledgments
S.Y. Haffert, R. Landman, and Y. Xin are supported by NWO award 184.036.004 and VI.Vidi.233.144. MagAO-X was developed with support from the NSF MRI Award No. 1625441. The Phase II upgrade program is made possible by the generous support of the Heising-Simons Foundation. MagAO-X uses the CACAO software package, which is supported by NSF Award No. 2410616. The GMagAO-X project is grateful for support from the University of Arizona Space Institute for the preliminary design phase and to the GMT for supporting final design.

\bibliography{report} 
\bibliographystyle{spiebib} 

\end{document}